\documentclass[letter,twocolumn]{jpsj2} 
\newcommand{\simle}
{\raisebox{-0.75ex}[-1.5ex]{$\;\stackrel{<}{\sim}\;$}}

\def\e{{\epsilon}}
\def\k{{ {\bf k} }}

\def\q{{ {\bf q} }}

\def\w{{\omega}}
\def\a{{\alpha}}

\title{
Electron-Phonon Mechanism for Superconductivity in 
Na$_{0.35}$CoO$_2$: \\
Valence-Band Suhl-Kondo Effect Driven by Shear Phonons
}

\author{
Keiji \textsc{Yada} and 
Hiroshi \textsc{Kontani}}

\inst{Department of Physics, Nagoya University,
Furo-cho, Nagoya 464-8602, Japan}

\abst{
To study the possible mechanism of superconductivity 
in Na$_{0.35}$CoO$_2$, we examine the interaction 
between all the relevant optical phonons (breathing and shear phonons)
and $t_{2g}$ $(a_{1g}+e_g')$ electrons of Co ions, and study the transition
temperature for $s$-wave superconductivity.
The obtained $T_{\rm c}$ is very low
when the  $e_g'$ valence bands are far below the Fermi level.
However, $T_{\rm c}$ is strongly enhanced when the top of the 
$e_g'$ valence bands is close to the Fermi level ({\it e.g.}, $-50$ meV),
thanks to the interband hopping of Cooper pairs caused by shear phonons.
This ``valence-band Suhl-Kondo mechanism'' due to shear phonons
is important in understanding the superconductivity in Na$_{0.35}$CoO$_2$.
By the same mechanism, the kink structure of the band dispersion
observed by angle-resolved photoemission spectroscopy (ARPES) measurements,
which indicates the strong mass enhancement 
($m^\ast/m\sim3$) due to optical phonons, is also explained.
}

\kword{Na$_x$CoO$_2$, electron-phonon coupling, valence-band Suhl-Kondo effect, shear phonon}

\begin{document}
\maketitle


Despite enormous theoretical and experimental effort,
the origin of superconductivity (SC) in Na$_{0.35}$CoO$_2\cdot y$H$_2$O
($T_{\rm c}=4.5$ K) is still controversial.
Several NMR/NQR measurements above $T_{\rm c}$
have revealed the existence of 
prominent magnetic fluctuations with finite momenta
 \cite{Sato,Ishida,Imai2},
which might remind someone of unconventional SC due to Coulomb interactions.
Recently, the anisotropy of Knight shift had been measured in aligned crystals
 \cite{Sato-NMR}
or in oriented powders
 \cite{Alloul,Ihara}.
Below $T_{\rm c}$, a sizable change in Knight shift is observed
for both ${\bf H}\perp {\bf c}$ and ${\bf H}\parallel {\bf c}$ in aligned crystals,
which indicates a singlet SC
 \cite{Sato-NMR}.
Note that no shift is observed for ${\bf H}\parallel {\bf c}$ in oriented powders
 \cite{Ihara}.

Significant information on the Fermi surface 
has been obtained by recent angle-resolved photoemission spectroscopy (ARPES) measurements:
They show that Na$_{0.35}$CoO$_2$($\cdot y$H$_2$O)
possesses a single hole like Fermi surface composed of
$a_{1g}$ orbitals in Co ions, whereas the other two bands
which originate from $e_{g'}$ orbitals are completely
below the Fermi level
 \cite{ARPES1,ARPES2,shimojima}.
They report that the top of $e_{g'}$-like valence bands is located
about $20\sim100$ meV below the Fermi level,
irrespective of the fact that a LDA band calculation predicts
the existence of Fermi surfaces of $e_{g'}$ bands,
which are composed of six small hole pockets just inside K points.
 \cite{Singh}.
Theoretically, the absence of small hole pockets
would be unfavorable for unconventional SC
 \cite{Mochizuki}.
It is noteworthy that the 
``weak pseudogap behavior'' observed in the uniform susceptibility 
below room temperature in Na$_{0.35}$CoO$_2$
  \cite{Sato}
is well reproduced by the fluctuation-exchange (FLEX) 
approximation only when small hole pockets are absent,
as observed in ARPES measurements
 \cite{Yada2}.


Besides the on-site Coulomb interaction, the existence of
a strong electron-phonon interaction (EPI) in Na$_{x}$CoO$_2$
is indicated by various optical measurements
 \cite{optical,Fano}.
In addition, the quasiparticle dispersion 
for the $a_{1g}$ band observed by ARPES measurement
has a prominent ``kink'' structure at around 70 meV
below the Fermi level, which is expected to originate
from phonons because the energy of optical phonons is also $\sim$ 70 meV.
The estimated mass-enhancement factor 
due to phonons is about three.
On the other hand,
the bandwidth of $t_{2g}$ bands observed by ARPES measurement
is around 0.8 eV, which is about half the bandwidth given 
in the LDA band calculation
 \cite{Singh}.
Thus, the mass-enhancement factor due to Coulomb interaction is about two
 \cite{Yada3}, 
so the total mass-enhancement factor becomes $2\times3=6$.
This fact would indicate that the EPI will be 
dominant in Na$_{x}$CoO$_2$.

\begin{figure}[htbp]
\begin{center}
\includegraphics[width=7cm]{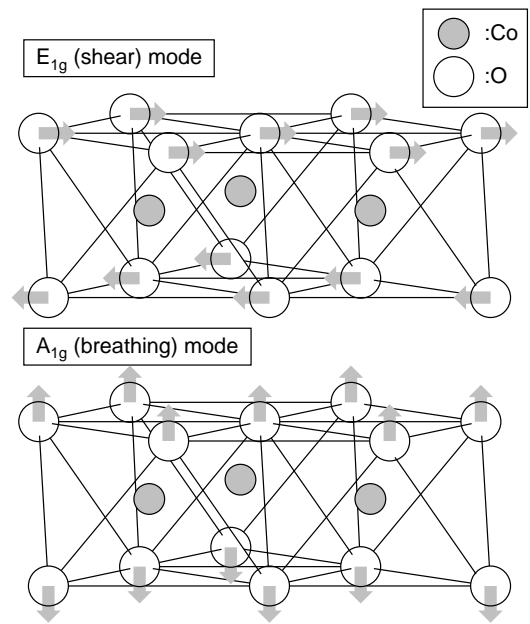}
\caption{Displacement of O ions 
by $E_{1g}$ (shear) mode and $A_{1g}$ (breathing) mode.}
\end{center}
\end{figure}

In the present work, we analyze the EPI
on the basis of the $d$-$p$ model for Na$_{0.35}$CoO$_2$
given in ref. 12, and
find that both shear and breathing optical phonons 
are strongly coupled with $t_{2g}$ electrons of Co ions.
If only the breathing phonon is taken into account,
the $T_{\rm c}$ for $s$-wave SC obtained 
by the Eliashberg equation is very low.
On the other hand, a much higher $T_{\rm c}$ is obtained
when we consider both shear and breathing phonons,
thanks to the interband hopping of Cooper pairs
between the $a_{1g}$ conduction band and the $e_{g}'$ valence bands.
This ``valence-band Suhl-Kondo (SK) mechanism'' due to shear phonons
also gives a large mass enhancement.

The structure of the CoO$_2$ layer is a triangular network of edge-sharing oxygen octahedra.
The trigonal deformation of the octahedra splits the $t_{2g}$ orbitals
into the $a_{1g}$ orbital and twofold degenerate $e_g'$ orbitals.
Here, we take the basis of $t_{2g}$ electrons as
$|a_{1g}\rangle=(|d_{xy}\rangle+|d_{yz}\rangle+|d_{zx}\rangle)/\sqrt 3$,
$|e'^1_{g}\rangle=(|d_{yz}\rangle-|d_{zx}\rangle)/\sqrt 2$,
and $|e'^2_{g}\rangle=(2|d_{xy}\rangle-|d_{yz}\rangle-|d_{zx}\rangle)/\sqrt 6$.
We introduce $V_t'$ to express the change of the crystalline 
electric splitting between $a_{1g}$ and $e_g'$ orbitals
from the value given by the LDA band calculation
 \cite{Yada2}.
In the point charge model, $V_t'$ {\it decreases} and $e_g'$-orbital level is raised
when the thickness of CoO$_2$ layer is reduced by trigonal distortion.

First, we focus on the zone center phonon modes.
NaCoO$_2$ has 7 kinds of irreducible representations for optical phonons per CoO$_2$ layer 
 \cite{linear}.
These phonons are separated into 
a soft group (3 modes) with frequencies below about 400 cm$^{-1}$
and a hard group (4 modes) with frequencies between 450 and 650 cm$^{-1}$.
The latter corresponds to the CoO$_2$ octahedron's oscillating modes,
which can give a strong EPI.
In these 4 modes, we find that
only $E_{1g}$ and $A_{1g}$ modes strongly couple with $t_{2g}$ electrons,
whereas the other two ungerade modes have no coupling with respect to the linear displacement of O ions.
In $E_{1g}$ and $A_{1g}$ modes,
O ions oscillate in the direction parallel and perpendicular to the CoO$_2$ layer, respectively, as shown in Fig. 1.
Hereafter, we call them "shear-mode phonons" and "breathing-mode phonons", respectively.
Note that the shear mode belongs to a two-dimensional representation.

Hereafter, we calculate the strength of the EPI via the frozen phonon method.
For this purpose, we ignore the effect of trigonal distortion for simplicity of calculation.
The displacement due to an $m$-mode phonon ($m=BR,SH1,SH2$)
is expressed as ${\bf u}_m=\sqrt{\frac{\hbar}{2M\omega_m}}(b_m+b^\dag_m)$,
where $M$ is the mass of an O ion, $\omega_m$ is the frequency of oscillation 
and $b^{(\dag)}_m$ is the annihilation (creation) operator for the phonons.
The Hamiltonian which denotes EPI for a multi orbital model is written as follows:
\begin{eqnarray}
H_{\rm EPI}=\frac{1}{\sqrt N}\hspace{-2mm}\sum^{BR,SH_{1,2}}_{m}\hspace{-1mm}\sum_{\k,\q,\sigma}\hat c^\dag_{{\bf k+q}\sigma}\hat V^m\hat c_{\k\sigma}(b_{m,{\bf q}}+b^\dag_{m,{\bf -q}}),
\end{eqnarray}
where $\hat c^{(\dag)}_{\k\sigma}=(c^{(\dag)}_{\k,\a_{1g},\sigma}\ c^{(\dag)}_{\k,\e'^1_{g},\sigma}\ c^{(\dag)}_{\k,\e'^2_{g},\sigma})$
is the annihilation (creation) operator's column (row) vector for electrons,
and $\hat V^m$ is the coupling for $m$-mode phonons.
Hereafter, we drop the $\q$-dependence of $\hat V^m$ and
use the matrices for zone center phonon for simplicity of calculation.

By choosing an appropriate coordinate axis for shear phonons,
the matrices have the following form
 \cite{preparation}:
\begin{eqnarray}
&&\hat V^{BR}=\left(
\begin{array}{ccc}
\ a_1\ &\ 0\ &\ 0\ \\
\ 0\ &\ a_2\ &\ 0\ \\
\ 0\ &\ 0\ &\ a_2\ 
\end{array}
\right),
 \label{eqn:VBR}
\end{eqnarray}
\begin{eqnarray}
&&\hat V^{SH^1}\hspace{-0.5em}=\hspace{-0.5em}\left(
\begin{array}{ccc}
0&b_1&0\\
b_1&0&-b_2\\
0&-b_2&0
\end{array}
\right),
\hat V^{SH^2}\hspace{-0.5em}=\hspace{-0.5em}\left(
\begin{array}{ccc}
0&0&-b_1\\
0&b_2&0\\
-b_1&0&-b_2
\end{array}
\right).
\nonumber\\
\end{eqnarray}
%
First, we consider the EPI originating from
the change of the Coulomb potential for $t_{2g}$ electrons due to the displacement of O ions, by considering O ions as point charges.
The obtained result is
$a^C_1=-C(1-\frac{4}{7}\frac{r^2_d}{a^2})$, $a^C_2=-C(1+\frac{2}{7}\frac{r^2_d}{a^2})$,
$b^C_1=C\frac{2}{7}\frac{r^2_d}{a^2}$, and $b^C_2=\frac{1}{4\sqrt 2}b^C_1$ on the order of $a^{-4}$
 \cite{preparation}.
Here, $a$ is the lattice constant between a Co site and O site, $r_d$ is the ionic radius of Co$^{3+}$,
and $C=\frac{2e^2}{a^2}\sqrt{\frac{\hbar}{2M\omega_m}}\cdot 2\sqrt{3}$;
$\sqrt{\frac{\hbar}{2M\omega_m}}=0.043$ \AA\ if we put $\omega= 60$ meV.
By using the values $a\approx 1.9$ \AA\ and $r_d \approx 0.61$ \AA\ , 
$C$ is estimated to be 0.87 eV.

Next, we consider the EPI originating from the change of the transfer integrals between Co and O.
According to Harrison's law ($t\propto a^{-4}$)
 \cite{Harrison},
$\delta t_{pd\pi}=-t_{pd\pi}\frac{4}{a}{\bf u}_m\cdot{\bf e}_{\rm O\mbox{-}Co}$,
where ${\bf e}_{\rm O\mbox{-}Co}=({\bf r}_{\rm O}-{\bf r}_{\rm Co})/a$.
The $2p$ orbitals of oxygen are filled with electrons.
When we consider the virtual process in which a hole of $t_{2g}$ orbitals transfers to $2p$ orbitals and turns back,
its energy shifts by $-\frac{t^2_{pd\pi}}{\Delta_{pd}}$ per O ion,
where $\Delta_{pd}$ is the charge transfer energy.
Thus, the variation in electron level accompanied by the change in $t_{pd\pi}$ is
$\delta\varepsilon_d=\frac{2t_{pd\pi}\delta t_{pd\pi}}{\Delta_{pd}}$.
Up to the fourth-order processes, we obtain $a^T_1=a^T_2=-2T$, and $b^T_1=T, b^T_2=T/\sqrt 2$,
where $T$ is given by
\begin{eqnarray}
&&\frac{16}{\sqrt 3}\frac{t^2_{pd\pi}}{\Delta_{pd}}\frac{1}{a}\sqrt{\frac{\hbar}{2M\omega_m}}
\Bigg\{1+\frac{1}{\Delta_{pd}}(|t_{pp\sigma}|+|t_{pp\pi}|)\nonumber\\
&&{}+\frac{1}{\Delta^2_{pd}}\left( \frac{17}{4}t^2_{pp\sigma}+\frac{29}{4}t^2_{pp\pi}+\frac{9}{2}|t_{pp\sigma}t_{pp\pi}| + \frac{t^2_{pd\pi}n_d}{6} \right)  \Bigg\}.
\nonumber
\end{eqnarray}
Here, we considered only nearest-neighbor hoppings between Co$-$O and O$-$O.
Using the values of Slater-Koster parameters given in ref. 12,
$T$ is estimated to be 0.158 eV.

Thus, we obtain $b_1=b^C_1+b^T_1=0.21$ eV, and $b_2=b^C_2+b^T_2=0.12$ eV for shear phonons.
As regards breathing phonons, we need to consider the chemical potential shift $\delta \mu$ caused by the $A_{1g}$ frozen phonon
since Tr\{$\hat V^{BR}$\} is finite as shown eq. (\ref{eqn:VBR}).
Because of the conservation of electron number,
$\rho_d(0)(\delta\varepsilon_d+\delta\mu)+\rho_p(0)\delta\mu=0$,
where $\rho_d(0)$ and $\rho_p(0)$ are the density of states (DOS) of $d$-electrons and $p$-electrons at the Fermi level.
Thus, the effective shift in the $3d$ level is
$\delta\tilde\varepsilon_d=\delta\varepsilon_d+\delta\mu=\frac{\rho_p(0)}{\rho_d(0)+\rho_p(0)}\delta\varepsilon_d$.
According to the band calculation
 \cite{Singh}
or in the present tight-binding model,
$\frac{\rho_p(0)}{\rho_d(0)+\rho_p(0)}\sim0.2$.
In conclusion, $a_1=0.2(a^C_1+a^T_1)\sim-0.2 \mbox{ eV} \approx a_2$.
Note that such a screening effect due to $\delta\mu$ is absent for shear phonons,
so the estimated values of $b_1$ and $b_2$ are reliable.

In what follows, we analyze the strong-coupling Eliashberg equation numerically.
According to first-principle linear response calculations
 \cite{linear},
the frequencies of shear and breathing phonons are about 500 cm$^{-1}$
and 600 cm$^{-1}$, respectively, and they are almost dispersionless.
To reduce the number of model parameters to simplify the analysis,
we take $\omega_D=550$ cm$^{-1}\sim68.2$ meV for both phonons.
In the present numerical calculation, we put
\begin{eqnarray}
a_1= 2a_2= -\alpha^{BR},
\ \
b_1= 4\sqrt 2 b_2= \alpha^{SH}.
\end{eqnarray}
We assume $\alpha^{BR}=\alpha^{SH}=0.2$, considering that the estimated values of $a_1$ and $b_1$ are about 0.2.
Both $a_1$ and $b_1$ are important parameters for the SC
because they are related to the $a_{1g}$ orbital, which makes the large Fermi surface around the $\Gamma$ point.
On the other hand, $a_2$ and $b_2$ are less important when small hole pockets composed of $e_g'$ orbitals are absent.
We use a one-loop approximation to calculate the self-energy $\Sigma(i\omega_n)$.
\begin{eqnarray}
&\displaystyle D(i\omega_n)=\frac{2\omega_D}{\omega^2_D+\omega^2_n},&\\
&\displaystyle \hat\Sigma(i\varepsilon_n)=\frac{T}{N}\sum_{\k,n',m}\hat V^m\hat G(\k,i\varepsilon_n-i\omega_{n'})\hat V^mD(i\omega_{n'}),\label{eq:sigma}&\\
&\hat G(\k,i\varepsilon_n)=\left(\left(\hat G^{(0)}(\k,i\varepsilon_n)\right)^{-1}-\hat\Sigma(i\varepsilon_n)\right)^{-1},&\label{eq:green}
\end{eqnarray}
where $\omega_n=2n\pi T$, and $\varepsilon_n=(2n+1)\pi T$.
These equations are solved self-consistently.

Figure 2 shows the $V_t'$ dependence of the inverse of the renormalization factor
$z^{-1}_\ell=1-\frac{\partial}{\partial\omega}\Sigma_\ell(\omega)\big |_{\omega\rightarrow0}$ at $T=0.005$ eV,
where $\Sigma_\ell$ is the normal self-energy for the $\ell$ orbital.
Here, $z^{-1}_\ell$ is the mass-enhancement factor for $\ell$-orbital
because the $\k$-dependence of the self-energy is absent in the present model.
We have verified that $z^{-1}_\ell$ is almost invariant against temperature in the present calculation.
At $(\alpha^{SH},\alpha^{BR})=(0.2, 0.2)$,
$z^{-1}_\ell$ for the $a_{1g}$ orbital decreases as $V_t'$ increases.
Note that the $e_{g}'$ band is completely
below the Fermi level for $V_t'>0.06$ eV.
To investigate the role of the two kinds of phonons,
we perform the calculation for $(\alpha^{SH},\alpha^{BR})=(0, 0.2)$ and $(\alpha^{SH},\alpha^{BR})=(0.2, 0)$ for comparison.
At $V_t'=0.06$ eV where the top of the $e_g'$ band is just at the Fermi level,
about 65\% of the mass enhancement for the $a_{1g}$ band is 
brought about by shear phonons, which mediate the interband process,
and this ratio gradually decreases as $V_t'$ increases.


\begin{figure}[htbp]
\begin{center}
\includegraphics[width=8cm]{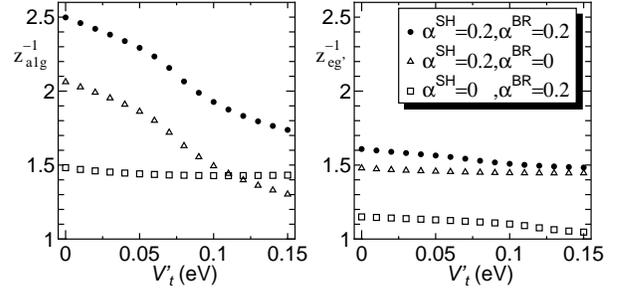}
\caption{Obtained $z^{-1}_\ell$
for $a_{1g}$ (left) and $e_g'$ (right) orbital.}
\end{center}
\end{figure}

Figure 3 is the renormalized band structure $E^*_\k$ for the $\Gamma$-K direction in the Brillouin zone obtained by
${\rm det}|\hat {G^R}^{-1}(E^*_\k)+\hat {G^A}^{-1}(E^*_\k)|=0$.
At $|E^*_\k|\simle\omega_D$,
the electronic band is renormalized significantly and
a kink structure appears in the $a_{1g}$ band.
This kink structure agrees well with the observation by ARPES measurements
 \cite{ARPES1}.
The $e'_g$ band is also renormalized by phonons, and its upper side has a flattened structure for $|E^*_\k|\simle\omega_D$.
The inset shows the top of the $e'_g$ band measured from the Fermi level, 
$\Delta$, 
which is different from the original value without interaction.
The small pockets disappear at $V'_t\simeq0.1$ eV without interaction
 \cite{Yada2},
but they disappear at $V'_t\simeq0.06$ eV when the EPI is taken into account.
This boundary value of $V'_t$ does not change
if only breathing phonons are taken into account,
which means that the shear phonons depress the $e'_g$ band to a lower energy
due to the off-diagonal elements of the self-energy.
Although there is a linear relationship between $\Delta$ and $V_t'$,
its gradient changes at $V'_t\simeq0.06$ eV, where the top of the $e'_g$ band just crosses the Fermi level.
We note that the second kink structure around 2$\omega_D$ is a famous artifact of dispersionless Einstein phonons,
so it will disappear when more realistic optical phonons with small dispersions are assumed.

\begin{figure}[htbp]
\begin{center}
\includegraphics[width=7cm]{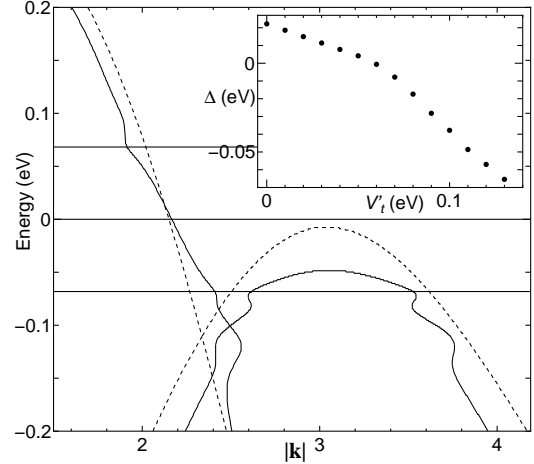}
\caption{Renormalized band structure $E^*_\k$ (solid line) for $\Gamma$-K direction and original band structure $\varepsilon_\k$ (dashed line) of tight-binding model at $V'_t=0.11$ eV.
The inset shows the $V'_t$ dependence of the top of the $e'_g$ band, $\Delta$.}
\end{center}
\end{figure}

Here, $T_{\rm c}$ is decided under the condition that the eigenvalue of the Eliashberg equation, $\lambda$, is equal to one:
\begin{eqnarray}
\lambda\hat \phi(i\varepsilon_n)&=&\frac{T}{N}\sum_{\k,n',m}\hat{V}^m\hat G(\k,i\varepsilon_{n'})\hat \phi(i\varepsilon_{n'})
 \nonumber \\ 
& &\times\hat G(-\k,-i\varepsilon_{n'})\hat{V}^m D(i\varepsilon_n-i\varepsilon_{n'}),
\end{eqnarray}
where we assumed a $\k$-independent gap function $\hat\phi(i\varepsilon_n)$.
Figure 4 shows the $V_t'$ dependence of $T_{\rm c}$.
At $(\alpha^{SH},\alpha^{BR})=(0.2, 0.2)$,
$T_{\rm c}$ increases as $V_t'$ decreases, whose change is much faster than that of the mass-renormalization factor.
The obtained $T_{\rm c}$ is high when the small pockets of the $e_g'$ band are present ($V_t'\simle0.06$ eV),
reflecting a huge DOS at the Fermi level.
It is noteworthy that $T_{\rm c}$ keeps high values for a while even after the small pockets disappear ($V_t'>0.06$ eV),
irrespective of the fact that the DOS suddenly decreases by one-third at $V_t'=0.06$ eV.
To find out the reason for this,
we study the cases of $(\alpha^{SH},\alpha^{BR})=(0, 0.2)$ and $(\alpha^{SH},\alpha^{BR})=(0.2, 0)$ for comparison.
In the former case, $T_{\rm c}$ is low and independent of $V_t'$.
In the latter case,
$T_{\rm c}$ drops quickly when the small pockets disappear
because shear phonons do not produce the attractive force 
between $a_{1g}$ electrons.
The reduction of $T_{\rm c}$ for $(\alpha^{SH},\alpha^{BR})=(0.2, 0.2)$ is slower than that for $(\alpha^{SH},\alpha^{BR})=(0.2, 0)$,
and $T_{\rm c}$ is considerably higher than that for $(\alpha^{SH},\alpha^{BR})=(0, 0.2)$ even when the small pockets disappear.
It is understood that the shear phonons
support the SC and raise $T_{\rm c}$ for a wide range of $V_t'$.
Because the shear phonons enable cooper pairs to transit from the $a_{1g}$ orbital to the $e_g'$ orbital,
a sort of SK mechanism works and $T_{\rm c}$ increases.
Indeed,
the gap function $\phi_\ell(\omega)$ is finite only for the $a_{1g}$ band for $(\alpha^{SH},\alpha^{BR})=(0, 0.2)$ for $V_t' >0.06$ eV;
in contrast, $\phi_\ell(\omega)$ is finite for both $a_{1g}$ and $e_g'$ bands when $(\alpha^{SH},\alpha^{BR})=(0.2, 0.2)$.
By the analytical study for 
$T_{\rm c}\ll -\Delta \ll \w_{\rm D}$
 \cite{preparation}, 
we derive the expression $T_{\rm c}\approx \w_{\rm D}\exp(-1/\lambda^*_{\rm eff})$,
where $\lambda^*_{\rm eff}$ is given by
\begin{eqnarray}
\lambda^*_{\rm eff}=\lambda^*_1+\frac{2\lambda^*_2\lambda^*_3\{\frac{1}{2}\log(\frac{\omega_D}{|\Delta|})+\frac{1}{\pi}\}}{1-\lambda^*_4\{\frac{1}{2}\log(\frac{\omega_D}{|\Delta|})+\frac{1}{\pi}\}},\label{eq:lambda}
\end{eqnarray}
where $\lambda^*_1=\frac{2a^2_1}{\omega_D}z_{a_{1g}}\rho_{a_{1g}}$, $\lambda^*_2=\frac{2b^2_1}{\omega_D}z_{a_{1g}}\rho_{a_{1g}}$,
$\lambda^*_3=\frac{2b^2_1}{\omega_D}z_{e_g'}\rho_{e_g'}$, and $\lambda^*_4=\frac{2a^2_2}{\omega_D}z_{e_g'}\rho_{e_g'}$.
In the derivation of eq. (\ref{eq:lambda}), we put $b_2=0$ and simplify the DOS as
$\rho_{a_{1g}}(\omega)=\rho_{a_{1g}}$, and 
$\rho_{e_g'}(\omega)=\rho_{e_g'}\theta(\Delta-\omega)$ per orbital.
When $|\Delta|\simle\omega_D$, the second term 
raises $\lambda^*_{\rm eff}$ and $T_{\rm c}$.
When $|\Delta|/\omega_D=1/4$ $(V_t'\simeq0.09)$, $\frac{1}{2}\log(\frac{\omega_D}{|\Delta|})+\frac{1}{\pi}\simeq 1$,
and we estimate that $\lambda^*_1=\lambda^*_2\approx0.25$, $\lambda^*_3\approx0.50$, and $\lambda^*_4\approx0.13$ in the present model.
Thus, $\lambda^*_{\rm eff}$ increases by 0.32 to a value of 0.57 owing to the second term,
which is recognized as the valence-band SK effect.

\begin{figure}[htbp]
\begin{center}
\includegraphics[width=7cm]{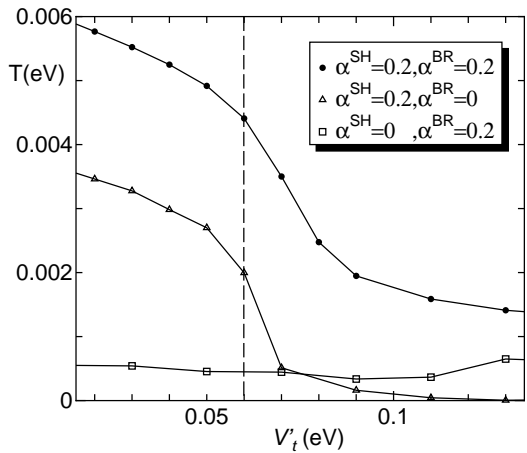}
\caption{$V_t'$ dependence of critical temperature $T_{\rm c}$. Small hole pockets are absent when $V_t'\ge0.06$.}
 \label{fig:Tc}
\end{center}
\end{figure}



Finally, we comment on the possible origin of the
anisotropy of the superconducting gap,
which is suggested by the absence of the coherence peak in $1/T_1T$
and by the specific heat measurement below $T_{\rm c}$.
We have shown that the gap function of $s$-wave SC due to EPI became highly anisotropic
when strong antiferromagnetic (AF) fluctuation coexist,
as realized in boron carbide, (Y,Lu)Ni$_2$B$_2$C 
 \cite{Kontani-boron}.
Thus, we expect that anisotropic $s$-wave state is realized in Na$_{0.35}$CoO$_2\cdot y$H$_2$O
because strong AF fluctuations are derived in the FLEX approximation
 \cite{Yada2}.
This is an important future problem.
We also comment on the effect of hydration on $T_{\rm c}$.
The hydration is expected to make $V_t'$ smaller since the thickness of CoO$_2$ layer is reduced,
which is actually suggested by NQR measurement
 \cite{Ishida-nyuQ}.
Figure \ref{fig:Tc} tells that the reduction of $V_t'$
due to hydration would explain why only hydrated samples show SC.
We stress that the $e_g'$ band in Na$_{0.35}$CoO$_2\cdot y$H$_2$O
does not intersect the Fermi level because the electronic heat coefficient
$\gamma$ is almost unchanged by hydration
 \cite{Sato}.
Otherwise, $\gamma$ should increase to more than three times 
the experimental value by hydration,
according to the change in the DOS at the Fermi level
 \cite{Yada2}.
In addition, the value of Knight shift is almost unchanged by hydration,
which also indicate the absence of small hole pockets in hydrated samples
 \cite{Alloul}.
Finally, we note that ref. 21 proposes a mechanism of $s$-wave SC due to the long-range Coulomb interactions.
A mechanism of $f$-wave SC due to EPI and Coulomb interaction is also proposed
 \cite{Greco}.

In summary,
we discussed the possibility of $s$-wave SC for Na$_x$CoO$_2$
based on a $d$-$p$ model with $a_{1g}$ and $e_g'$ bands, 
by considering EPI for breathing and shear phonons.
The obtained $T_{\rm c}$ is strongly enhanced 
when shear phonons are taken into account,
in addition to the breathing phonons.
The estimated EPI for both phonons
are large enough to realize $s$-wave SC
against the strong Coulomb interaction in Na$_x$CoO$_2$.
We find that $T_{\rm c}$ is still high even if
small hole pockets are absent ($V_t\ge0.6$ eV),
although the DOS at the Fermi level suddenly decreases
to one-third of its value at $V_t\ge0.6$ eV.
This enhancement of $T_{\rm c}$ is brought about by the 
valence-band SK mechanism due to shear phonons.
In the future, we will study the effect of Coulomb interaction
to obtain a realistic $T_{\rm c}$, and to explain the
anisotropy of the superconducting gap.

We thank M. Sato, Y. Kobayashi, K. Ishida, Y. Ihara, Y. Matsuda, T. Sato and T. Shimojima
for valuable discussions on experiments.
We also thank K. Yamada, D. S. Hirashima, Y. Tanaka,
K. Kuroki and the authors of ref. 16 for useful comments 
and discussions.



\begin{thebibliography}{99}


\bibitem{Sato}
 M. Yokoi, T. Moyoshi, Y. Kobayashi, M. Soda,
 Y. Yasui, M. Sato and K. Kakurai:
 J. Phys. Soc. Jpn. {\bf 74} (2005) 3046.

\bibitem{Ishida}
 Y. Ihara, K. Ishida, K. Yoshimura, K. Takada, 
 T. Sasaki, H. Sakurai and E. T. Muromachi:
 J. Phys. Soc. Jpn. {\bf 74} (2005) 2177.

\bibitem{Imai2}
 F. L. Ning and T. Imai:
 Phys. Rev. Lett. {\bf 94} (2005) 227004.

\bibitem{Sato-NMR}
 Y. Kobayashi, H. Watanabe, M. Yokoi, T. Moyoshi, Y. Mori and M. Sato:
 J. Phys. Soc. Jpn. {\bf 74} (2005) 1800.

\bibitem{Alloul}
 I. R. Mukhamedshin, H. Alloul, G. Collin and N. Blanchard:
 Phys. Rev. Lett. {\bf 94} (2005) 247602.

\bibitem{Ihara}
 Y. Ihara, K. Ishida, H. Takeya, C. Michioka, M. Kato, Y. Itoh, 
 K. Yoshimura, K. Takada, T. Sasaki, H. Sakurai and E. T. Muromachi:
 J. Phys. Soc. Jpn. {\bf 75} (2006) 013708.



\bibitem{ARPES1}
 H.-B. Yang, Z.-H. Pan, A. K. P. Sekharan, T. Sato, S. Souma, 
 T. Takahashi, R. Jin, B. C. Sales, D. Mandrus, A. V. Fedorov, 
 Z. Wang and H. Ding:
 Phys. Rev. Lett. {\bf 95} (2005) 146401.
\bibitem{ARPES2}
 M. Z. Hasan, D. Qian, Y. Li, A. V. Fedorov, Y.-D. Chuang, 
 A. P. Kuprin, M. L. Foo and R. J. Cava: cond-mat/0501530.
\bibitem{shimojima}
 T. Shimojima {\em et al}: unpublished.
\bibitem{Singh}
 D. J. Singh: Phys. Rev. B {\bf 61} (2000) 13397.

\bibitem{Mochizuki}
 M. Mochizuki, Y. Yanase and M. Ogata:
 J. Phys. Soc. Jpn. {\bf 74} (2005) 1670.




\bibitem{Yada2}
 K. Yada and H. Kontani:
 J. Phys. Soc. Jpn. {\bf 74} (2005) 2161.

\bibitem{optical}
 D. Wu, J. L. Luo and N. L. Wang:
 Phys. Rev. B {\bf 73} (2006) 014523.

\bibitem{Fano}
 S. Lupi, M. Ortolani and P. Calvani:
 Phys. Rev. B {\bf 69} (2004) 180506(R).

\bibitem{Yada3}
 K. Yada and H. Kontani:
 cond-mat/0507066.



\bibitem{linear}
 Z. Li, J. Yang, J. G. Hou and Q. Zhu:
 Phys. Rev. B {\bf 70} (2004) 144518.


\bibitem{preparation}
 K. Yada and H. Kontani:
 in preparation.

\bibitem{Harrison}
 W. A. Harrison:
 {\it Elementary Electronic Structure}
 (World Scientific, Singapore, 1999).


\bibitem{Kontani-boron}
 H. Kontani:
 Phys. Rev. B {\bf 70} (2004) 054507.

\bibitem{Ishida-nyuQ}
 Y. Ihara, K. Ishida, C. Michioka, M. Kato, K. Yoshimura, 
 K. Takada, T. Sasaki, H. Sakurai and E. T. Muromachi:
 J. Phys. Soc. Jpn. {\bf 74} (2005) 867.

\bibitem{kuroki}
 K. Kuroki, S. Onari, Y. Tanaka, R. Arita and T. Nojima: cond-mat/0508482

\bibitem{Greco}
 A. Foussats, A. Greco, M. Bejas and A. Muramatsu:
 Phys. Rev. B {\bf 72} (2005) 020504(R)
\end{thebibliography}
\end{document}